\documentclass[twocolumn]{svjour3}          

\usepackage[latin1]{inputenc}
\usepackage{times}
\usepackage{amsmath}
\usepackage{amssymb}
\usepackage{subfigure}
\usepackage{mathrsfs}
\usepackage{graphicx}
\usepackage{enumerate}
\usepackage[compress]{cite}
\usepackage{microtype}
\usepackage{float}

\smartqed  
\begin{document}

\title{Performance Analysis and Optimisation of FFR-Aided OFDMA Networks using Channel-Aware Scheduling}


\author{Jan Garc\'ia-Morales         \and
        Guillem~Femenias             \and 
        Felip Riera-Palou
}


\institute{    Mobile Communications Group -- University of the Balearic Islands (UIB) -- Palma 07122, Illes Balears (Spain)\\
              \email{\{jan.garcia,guillem.femenias,felip.riera\}@uib.es}           
}

\date{Received: date / Accepted: date}

\maketitle

\begin{abstract}
Modern cellular standards typically incorporate interference coordination schemes allowing near universal frequency reuse while preserving reasonably high spectral efficiencies over the whole coverage area. In particular, fractional frequency reuse (FFR) and its variants are deemed to play a fundamental role in the next generation of cellular deployments (B4G/5G systems). This paper presents an analytical framework allowing the downlink performance evaluation of FFR-aided OFDMA networks when using channel-aware scheduling policies. Remarkably, the framework contemplates the use of different rate allocation strategies, thus allowing to assess the network behaviour under ideal (capacity-based) or realistic (throughput-based) conditions. Analytical performance results are used to optimise the FFR parameters as a function of, for instance, the resource block scheduling policy or the density of UEs per cell. Furthermore, different optimisation designs of the FFR component are proposed that allow a tradeoff between throughput performance and fairness by suitably dimensioning the FFR-defined cell-centre and cell-edge areas and the corresponding frequency allocation to each region. Numerical results serve to confirm the accuracy of the proposed analytical model while providing insight on how the different parameters and designs affect network performance.
\end{abstract}


\section{Introduction}\label{sec:Introduction}


Orthogonal frequency division multiple access (OFDMA) has been adopted as the downlink multiple access scheme for state-of-the-art cellular communications standards such as Long-Term Evolution (LTE) and LTE-Advanced (LTE-A) \cite{Dahlman13}. In OFDMA, a wideband frequency-selective fading channel is decomposed into a set of orthogonal narrow-band subchannels. These subchannels are jointly used with a time-slotted frame pattern to provide a set of frequency/time resources (also known as resource blocks (RBs)), which are distributed among cells based on predefined frequency reuse plans. The orthogonality among RBs makes the intra-cell interference negligible. However, the use of \emph{aggressive} high spectral efficiency universal frequency reuse plans, with all cells using the same set of RBs, cause the OFDMA-based networks to suffer from very high levels of inter-cell interference (ICI), particularly affecting the UEs located in the cell-edge areas. With the aim of mitigating ICI experienced by the cell-edge users while still achieving high spectral efficiencies, a myriad of ICI control (ICIC) strategies have been proposed \cite{Hamza13}, among which \emph{static} fractional frequency reuse (FFR) and all its variants show a good tradeoff between cell-edge throughput enhancement, provision of high spectral efficiency and implementation complexity \cite{Saquib13}.

In the downlink of FFR-aided OFDMA-based networks, the time/frequency channel quality of the set of RBs allocated to a given base station (BS), typically measured in terms of the signal-to-interference-plus-noise ratio (SINR), varies for different UEs. Such variations in channel conditions can be exploited by using channel-aware schedulers able to allocate each RB to a UE with favourable channel conditions at a given scheduling time slot. Opportunistic maximum SINR (MSINR) scheduler \cite{Knopp95} makes the most of the multiuser diversity by allocating the RBs to the UEs experiencing the best channel conditions. Although applying this scheduling rule aims at maximising the spectral efficiency of the system, it raises a serious fairness problem, with UEs suffering from bad channel conditions over extended periods of time experiencing a dramatic quality-of-service (QoS) degradation. In order to provide a reasonable tradeoff between spectral efficiency and fairness, a proportional fair (PF) scheduling rule was proposed by Kelly \emph{et al.} in \cite{Kelly98} and further developed by Shakkottai \emph{et al.} in \cite{Shakkottai00}. In this case, scheduling decisions rely on a weighted version of the instantaneous channel behaviour, with the weighting coefficient for a given UE being inversely proportional to the average channel behaviour during the time this UE has gained access for transmission. In this way, the RBs end up being allocated to UEs experiencing the relatively best channel conditions in comparison to their average channel state and thus, using a PF scheduler, the possibility of a UE with a very bad link suffering from long periods of starvation is drastically reduced.


The problem of analytically evaluating the performance of FFR-aided OFDMA-based cellular networks has been recently tackled using tools drawn from the theory of stochastic geometry, where the BSs are distributed using Poisson Point Processes (PPPs) \cite{Novlan11} (see also \cite{ElSawy13} and references therein). Stochastic geometry-based frameworks analyze the system performance by spatially averaging over all possible network realisations. Using this approach allows characterising the performance for an entire network but precludes from accurately analysing the performance of a given cell. In particular, it is a metric of practical importance to network designers that, provided a planned set of BS locations along with traffic load conditions, may be interested in calculating the performance obtained within a specific region in the coverage area of the network. In order to overcome the limitations of the stochastic geometry approach, Heath et al. \cite{Heath13} proposed a rather involved upper bound to model the interference by inscribing a circular cell within the weighted Voronoi cell provided by the PPP-based model. Nonetheless, characterising the cellular layouts using stochastic geometry makes it extremely difficult, if at all possible, to accurately model the use of the ICIC techniques proposed for modern well-planned macrocell networks, including FFR and its variants and/or BS cooperation schemes.

In contrast to the above background work, which relies on the use of stochastic geometry to model the cellular environment, Fan Jin \emph{et al.} \cite{Jin13} consider an FFR-aided twin-layer OFDMA network where stochastic geometry is used to characterise the random distribution of femtocells, and the macrocells are overlaid on top of the femtocells following a regular tessellation. One of the main limitations of this work, however, is the use of rather unrealistic assumptions such as neglecting the small scale fading effects and, consequently, limiting the proposed analytical framework to resource allocation schemes based on the round robin (RR) scheduling policy. Similar approaches, lacking the consideration of small scale fading and scheduling policies, are also proposed by Assaad in \cite{Assaad08} and Najjar \emph{et al.} in \cite{Najjar09} to optimise FFR-based parameters in a single-tier network. These limitations have been overcome in part in our recent contribution \cite{Garcia-Morales15} (see also \cite{Femenias15}), but only taking into account the use of opportunistic MSINR schedulers and capacity-based resource allocation strategies, and thus neglecting the use of more practical schedulers such as PF and realistic adaptive modulation and coding (AMC) schemes.


Despite the attractive properties of the PF scheduler, it makes the spectral efficiency analysis of the system rather involved \cite{Choi07,Liu10,Wu11,Parruca13}. In this paper we present a novel analytical framework allowing the throughput performance evaluation of an FFR-aided OFDMA-based cellular network using a PF scheduling policy. The proposed analytical framework turns out to extend the applicability of the analysis to the optimisation of the FFR in the spatial and frequency domains using different optimisation criteria, namely, the fixed-spectrum-allocation-factor design, the area-proportional design or the QoS-constrained design.  Furthermore, and departing from previous works, the framework introduced here can accommodate different rate allocation strategies, such as, continuous rate allocation (CRA), useful to examine performance upper bounds, or discrete rate allocation (DRA), suitable to examine the performance of systems based on AMC. We further demonstrate how the performance of both RR and MSINR scheduling rules can be derived as special cases of the PF scheduler. Although results are obtained for an FFR-aided deployment, this analytical framework opens the door to the theoretical spectral efficiency evaluation of OFDMA-based cellular networks using more sophisticated ICIC techniques such as soft frequency reuse, adaptive frequency reuse or network MIMO \cite{Pastor15a}, as well as to the assessment of cellular multi-tier networks where the macro-cellular network is underlaid by different tiers of pico- and femto-cellular BSs \cite{Garcia-Morales15}.


The rest of the paper is organised as follows. In Section \ref{sec:System_model} the system model under consideration is introduced alongside with the key simplifying assumptions made for the sake of analytical tractability. Section \ref{sec:Macro_analysis} elaborates on the analytical framework used to characterise the average throughput performance of the FFR-aided OFDMA-based cellular network using a PF scheduling rule. The framework is then extended to both the MSINR and RR scheduling strategies. Optimal designs for these set-ups are presented in Section \ref{sec:Optimisation}. Extensive analytical and simulation results are provided in Section \ref{sec:Numerical_results}. Finally, the main outcomes of this work are recapped in Section \ref{sec:Conclusion}.

\section{System model}\label{sec:System_model}
\begin{figure}
\centering
\includegraphics[width=0.45\textwidth]{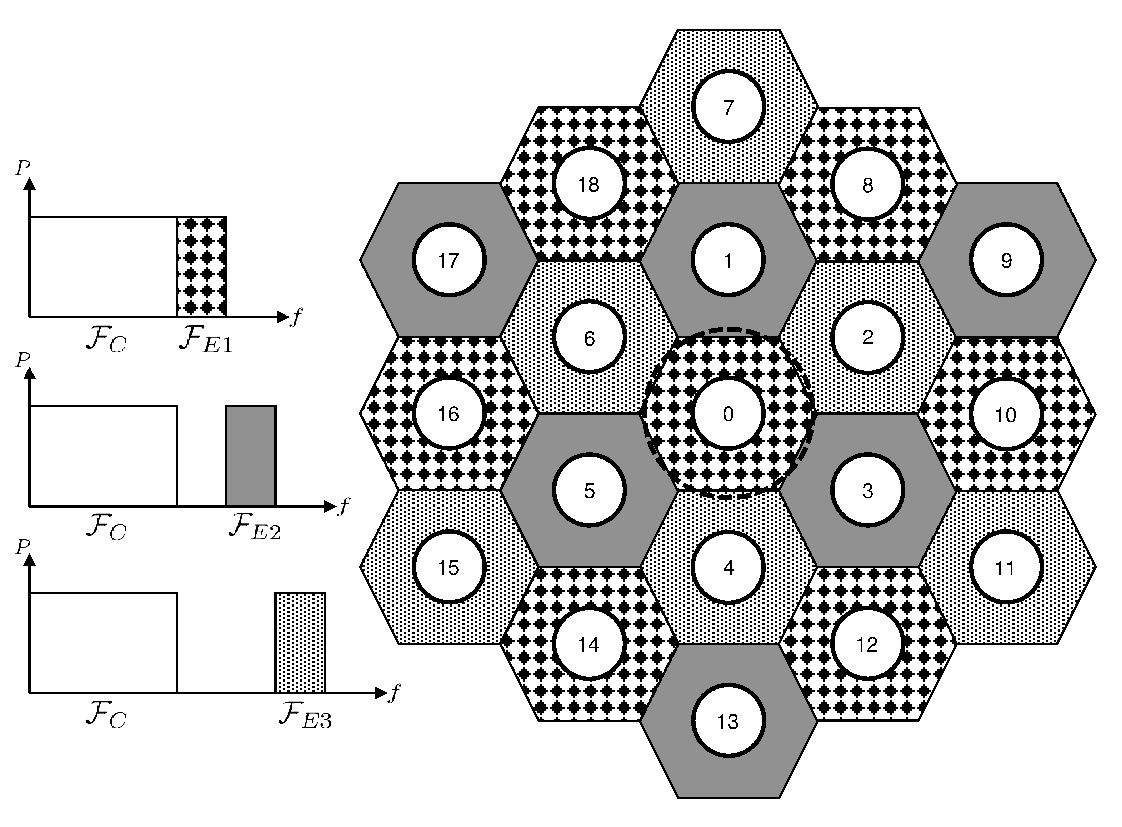}
\caption{Topology of the FFR-aided OFDMA-based two-tier network.}
\label{fig:Network_model}
\end{figure}
\subsection{Network topology model}
Let us consider the downlink of an FFR-aided OFDMA-based cellular system where a set of macrocellular BSs are assumed to be deployed following a conscious planning and thus, are regularly arranged over the whole coverage area. This cellular environment can be safely modeled as a regular tessellation of hexagonally-shaped coverage areas, as shown in Fig. \ref{fig:Network_model}, with the BSs located at the centre of the hexagons\footnote{Omnidirectional antenna BSs are assumed in this paper. In future work this will be extended to consider the use of sectorisation.}. For the sake of analytical tractability, the central cell, covered by BS 0, which will be referred to as the tagged BS or the BS of interest, will be approximated by a circle whose area is the same as the hexagonal one (see Fig. \ref{fig:Network_model}). That is, assuming that the side of the regular hexagon is $R_h$, the radius of the circular cell is $R_m=R_h\sqrt{3\sqrt{3}/(2\pi)}$.

The locations of the UEs at a given time instant are assumed to form a stationary PPP of normalised intensity $\lambda_m$ (measured in UEs per area unit). The PPP assumption is valid when the users are independently and uniformly distributed over the service coverage area, which is a reasonable approach to obtain analytical expressions for the spectral efficiency of the network. A consequence of this assumption is that the probability distribution of the number $M_\mathcal{S}$ of users falling within any spatial region $\mathcal{S}$ of area $A_\mathcal{S}$ follows a Poisson distribution, thus implying
\begin{equation}\label{e0}
Pr\{M_\mathcal{S}=k\} = \frac{(\lambda_m A_\mathcal{S})^k e^{-\lambda_m A_\mathcal{S}}}{k!}.
\end{equation}

In order to control the ICI, UEs are classified according to the received average SINR as either cell-centre users, when the received average SINR is above a given threshold, or cell-edge users, when it is below the threshold. An FFR scheme is then applied by allocating non-overlapping RBs to cell-centre and cell-edge users, while employing a frequency reuse factor equal to one for the cell-centre users and a higher frequency reuse factor for the cell-edge users that, without loss of generality, is assumed to be three in this paper. Again, for analytical tractability, cell-centre and cell-edge regions will be separated by a circumference of radius $R_{th}$ (threshold distance or threshold radius).

As previously stated, transmission between the BS and active MSs is organized in time-frequency resource allocation units that are known as RBs. Each RB is formed by a slot in the time axis and a subband in the frequency axis:
\begin{itemize}
   \item In the time axis, each RB occupies a time-slot of a fixed duration $T_s$, assumed to be less than the channel coherence time. Thus, the channel fading can be considered constant over the whole slot and it only varies from slot to slot, i.e., a slot-based block fading channel is assumed. Each of these slots consists of a fixed number $N_o$ of orthogonal frequency-division multiplexing (OFDM) symbols of duration $T_o + T_{CP} = T_s/N_o$, where $T_{CP}$ is the cyclic prefix (CP) duration.

   \item Slotted transmissions take place over a bandwidth $B$, which is divided into $N_f$ orthogonal subcarriers, out of which $N_d$ are used to transmit data and $N_p$ are used to transmit pilots and to set guard frequency bands. The $N_d$ data subcarriers are divided into $N_b$ orthogonal subbands, indexed by the set $\mathcal{F}_T$, each consisting of $N_{sc}$ adjacent subcarriers and with a bandwidth $B_b = B N_{sc}/N_f$ small enough to assume that all subcarriers in a subband experience frequency flat fading.
\end{itemize}
Thus, during each scheduling period of duration $T_s$, the total system bandwidth $B$ is exploited by means of a set $\mathcal{F}_T$ of RBs (or subbands), which is split into a set $\mathcal{F}_C$ of RBs allocated to the cell-centre and a set $\mathcal{F}_T \backslash \mathcal{F}_C$ of RBs allocated to the cell-edge. The set $\mathcal{F}_T \backslash \mathcal{F}_C$ is further split into three equal parts, namely $\mathcal{F}_{E1}$, $\mathcal{F}_{E2}$ and $\mathcal{F}_{E3}$, of size $|\mathcal{F}_E|$, which are allocated to cell-edge UEs in such a way that adjacent cells will operate on different sets of RBs, as shown in Fig.~\ref{fig:Network_model}. Note that, denoting by $N_C$ and $N_E$ the number of RBs allocated to the cell-centre area and each of the cell-edge regions, respectively, we have that $N_b=N_C+3 N_E$.

\subsection{Channel model}
The downlink channel is subject to path loss and small-scale fading\footnote{In line with Xu \emph{et al.} \cite{Xu12} and Jin \emph{et al.} \cite{Jin13}, note that, for the sake of analytical simplicity, only pathloss and small scale fading are considered in this paper. In future work this will be extended to consider large scale fading (shadowing) as well.}. The path loss characterising the link between the $b$th BS and the $u$th UE can be modeled as (in dB)
\begin{equation}\label{e1}
L\left(d_{b,u}\right) = K +10\alpha \log_{10}\left(d_{b,u}\right),
\end{equation}
where $K$ and $\alpha$ are, respectively, a fixed path loss and the path loss exponent, and $d_{b,u}$ is the distance (in metres) between the UE and the BS.

The instantaneous SINR experienced by UE $u$ in the cell of interest on any of the $N_{sc}$ subcarriers conforming the $n$th RB during the scheduling period $t$ can be expressed as
\begin{equation}\label{e2}
   \gamma_{u,n}(t) = \frac{ P_c L(d_{0,u})|H_{0,u,n}(t)|^2}{N_0 \Delta f + I_{u,n}(t)},
\end{equation}
where, assuming the use of uniform power allocation,
\begin{equation}
   P_c=\frac{P_T}{N_{sc}(|\mathcal{F}_C|+|\mathcal{F}_E|)}
\end{equation}
is the power allocated per subcarrier, with $P_T$ denoting the available transmit power at the BS, $H_{b,u,n}(t)\sim \mathcal{C N} (0,1)$ is the frequency response resulting from the small-scale fading channel linking the $b$th BS to UE $u$ on the $n$th RB during scheduling period $t$, $N_0$ is the noise power spectral density, $\Delta f=B_b/N_{sc}$ is the subcarrier bandwidth, and $I_{u,n}(t)$ denotes the interference term that is given by
\begin{equation}\label{e3}
   I_{u,n}(t)=\sum_{b\in \Phi_n} P_c L\left(d_{b,u}\right)|H_{b,u,n}(t)|^2,
\end{equation}
with $\Phi_n$ representing the set of interfering BSs, which is RB-dependent as, because of the use of FFR, the set of interfering BSs for RBs allocated to the cell-centre is different from that for RBs allocated to the cell-edge. In fact,
\begin{equation}\label{e4}
  \Phi_n = \begin{cases}
              \{1,2,...,18\}, & n\in \mathcal{F}_C,\\
              \{8,10,12,14,16,18\}, & n\in \mathcal{F}_E.
           \end{cases}
\end{equation}
As an important notational remark, note that $L\left(d_{b,u}\right)$ can be expressed in terms of UEs' polar positions with respect to BS 0 as $L(d_{0,u},\theta_{0,u})$ and thus, strictly speaking, $\gamma_{u,n}(t)$ is a function of $d_{0,u}$ and $\theta_{0,u}$. Nevertheless, it is shown in \cite{Maqbool10} that the instantaneous SINR in multi-cell networks barely depends on the polar angle and thus, from this point onwards, the dependence of $\gamma_{u,n}(t)$ on $\theta_{0,u}$ will be omitted.

\subsection{Rate allocation}

In the downlink of multi-rate systems using AMC, an SINR estimate is obtained at the receiver of each UE and it is then fed back to the BS so that the transmission mode, comprising a modulation format and a channel code, can be adapted in accordance to the instantaneous channel characteristics\footnote{The decision on which is the most adequate transmission mode to be used in the next scheduling interval can also be taken at the receiver side and then be fed back to the transmitter.}. If UE $u$ is allocated RB $n$ over time slot $t$, then the BS selects a modulation and coding scheme (MCS) that can be characterised by a transmission rate $\rho\left(\gamma_{u,n}(t)\right)$ (measured in bits per second) on each subcarrier of the RB. As each RB contains $N_{sc}$ subcarriers, the aggregated data rate allocated to UE $u$ on RB $n$ during time slot $t$ will be given by
\begin{equation}
\label{eq:rm}
   r\left(\gamma_{u,n}(t)\right) = N_{sc} \rho\left(\gamma_{u,n}(t)\right).
\end{equation}
In real systems there is a non negligible probability that a block error occurs during transmission, hence, the effective data rate allocated to UE $u$ on RB $n$ during time slot $t$ can be expressed as
\begin{equation}
\begin{split}
   \tilde{r}\left(\gamma_{u,n}(t)\right)&=\vartheta\left(\gamma_{u,n}(t)\right) r\left(\gamma_{u,n}(t)\right) \\
                                        &=N_{sc}\vartheta\left(\gamma_{u,n}(t)\right) \rho\left(\gamma_{u,n}(t)\right),
\end{split}
\end{equation}
where
\begin{equation}
   \vartheta\left(\gamma_{u,n}(t)\right)=\begin{cases}
                                            1 & \text{successful transmission of RB} \\
                                            0 & \text{otherwise.}
                                         \end{cases}
\end{equation}

\paragraph{Discrete-rate allocation (DRA)}: AMC strategies use a discrete set $\mathcal{N}_m=\{0,1,\ldots,N_m\}$ of MCSs, and each MCS is characterised by a particular transmission rate $\varrho^{(m)}$, with $\varrho^{(1)}<\ldots<\varrho^{N_m}$. The data rate $\varrho^{(0)}=0$ corresponds to the no-transmission mode, that is, the mode selected when the channel is so bad that no bits can be transmitted to the corresponding UE while guaranteeing the prescribed QoS constraints.

\begin{figure}
\centering
\includegraphics[width=\linewidth]{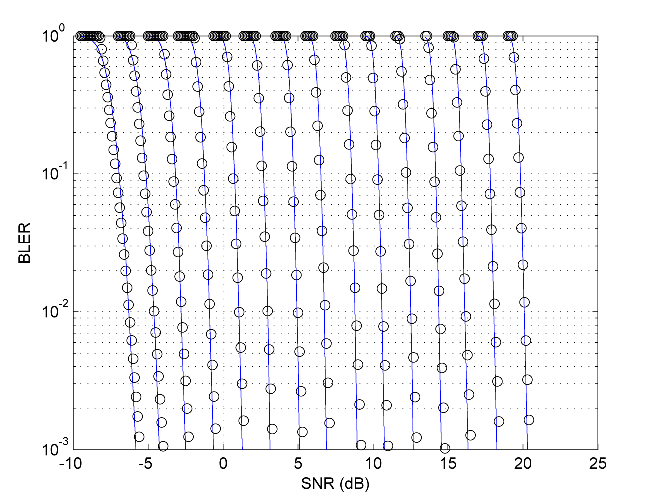}
\caption{Simulated (solid lines, obtained from \cite{Ikuno13}) \emph{vs} fitted (markers) BLER values for the different LTE/LTE-A modes.}
\label{fig:BLER_LTE}
\end{figure}

Transmission rate $\varrho^{(m)}$ can be related to block error rate (BLER) observed by MS $u$ and instantaneous SINR $\gamma_{u,n}(t)$ as \cite[Chapter 9]{Goldsmith05} (see also \cite{wang2007unified})
\begin{equation}
\label{eq:block-error-rate-relation}
\begin{split}
	&\epsilon(\gamma_{u,n}(t),\varrho^{(m)}) \\
    &\quad= \begin{cases}
                1, & \gamma_{u,n}(t) < \gamma^{(m)} \\
                \kappa_1^{(m)} \exp\left(-\kappa_2^{(m)} \gamma_{u,n}(t)\right), & \text{otherwise,}
            \end{cases}
\end{split}
\end{equation}
where $\kappa_1^{(m)}$ and $\kappa_2^{(m)}$ are modulation- and code-specific constants that can be accurately approximated by exponential curve fitting. For instance, in the particular case of LTE/LTE-A, any UE $u$ has $N_m=15$ transmission modes available plus the no transmission mode, allowing the number of coded bits per symbol to vary between $\varrho^{(1)} = 0.15$ bits/symbol (4-QAM and 0.08 coding rate) and $\varrho^{(15)} = 5.55$ bits/symbol (64-QAM and 0.93 coding rate). Figure~\ref{fig:BLER_LTE} shows the resulting BLER performance of the different transmission modes on an equivalent AWGN channel \cite{Ikuno13}, with the corresponding approximation obtained by least-squares fitting using \eqref{eq:block-error-rate-relation}. The leftmost curve corresponds to the lowest (most robust) transmission mode ($\varrho^{(1)}$) and each successive curve corresponds to the immediately higher transmission mode. Table~\ref{tab:BLER_LTE} presents the fitting parameters ($\kappa_1^{(m)}$, $\kappa_2^{(m)}$ and $\gamma^{(m)}$) for the different LTE/LTE-A transmission modes.

\begin{table}
\centering
\renewcommand{\arraystretch}{1.2}
\caption{BLER fitting parameters for the different LTE/LTE-A modes.}
\scalebox{0.9}{
\begin{tabular}{|c|c|c|c|c|}  \hline
  Mode ($m$) & $\varrho^{(m)}$ (bits/symbol)& $\kappa_1^{(m)}$ & $\kappa_2^{(m)}$ & $\gamma^{(m)}$ (dB)\\ \hline
  1 & 0.15 & 3.270e+03 & 53.678 & -8.217 \\
  2 & 0.23 & 4.067e+04 & 43.868 & -6.163 \\
  4 & 0.60 & 2.677e+07 & 26.549 & -1.910 \\
  5 & 0.88 & 1.005e+09 & 19.785 & 0.202 \\
  6 & 1.18 & 9.340e+09 & 13.888 & 2.183 \\
  7 & 1.48 & 1.160e+10 & 8.790 & 4.210 \\
  8 & 1.91 & 1.443e+11 & 6.358 & 6.065 \\
  9 & 2.41 & 1.000e+10 & 3.590 & 8.071 \\
  10 & 2.73 & 1.000e+10 & 2.373 & 9.869 \\
  11 & 3.32 & 1.000e+10 & 1.526 & 11.790 \\
  12 & 3.90 & 1.000e+10 & 0.991 & 13.663 \\
  13 & 4.52 & 1.000e+10 & 0.665 & 15.396 \\
  14 & 5.12 & 1.000e+10 & 0.426 & 17.329 \\
  15 & 5.55 & 1.000e+10 & 0.268 & 19.333 \\  \hline
\end{tabular}}
\label{tab:BLER_LTE}
\end{table}

Using the instantaneous SINR $\gamma_{u,n}(t)$, which has been fed back from the receiver, and assuming a maximum allowable BLER $\check{\epsilon}$, the transmitter can employ \eqref{eq:block-error-rate-relation} to select the most adequate MCS scheme as the one with transmission rate
\begin{equation}
\label{eq:dicrete-rate-amc-ideal_CSIT}
   \rho\left(\gamma_{u,n}(t)\right) = \max \left\{ \varrho^{(m)} \ : \ \epsilon \left(\gamma_{u,n}(t),\varrho^{(m)}\right) \leq \check{\epsilon}\right\}.
\end{equation}
In this case, the potential transmission rate $\rho\left(\gamma_{u,n}(t)\right)$ and BLER $\epsilon\left(\gamma_{u,n}(t)\right)$ can be expressed, respectively, using the staircase functions
\begin{equation}
\label{eq:dicrete-rate-amc-staircase-ideal-CSIT}
   \rho\left(\gamma_{u,n}(t)\right) =     \begin{cases}
                                              \varrho^{(0)},  & 0 \leq \gamma_{u,n}(t) < \Gamma^{(1)} \\
                                              \varrho^{(1)}, & \Gamma^{(1)} \leq \gamma_{u,n}(t) < \Gamma^{(2)} \\
                                                             & \qquad \vdots \\
                                              \varrho^{(N_m)}, & \Gamma^{(N_m)} \leq \gamma_{u,n}(t) < \infty
                                          \end{cases}
\end{equation}
and
\begin{equation}
\begin{split}
\label{eq:dicrete-rate-amc-BLER-staircase-ideal-CSIT}
   &\epsilon\left(\gamma_{u,n}(t)\right) \\
   &\     = \begin{cases}
               \kappa_1^{(1)} e^{-\kappa_2^{(1)} \gamma_{u,n}(t)}, & \Gamma^{(1)} \leq \gamma_{u,n}(t) < \Gamma^{(2)} \\
                              & \qquad \vdots \\
               \kappa_1^{(N_m)} e^{-\kappa_2^{(N_m)} \gamma_{u,n}(t)}, & \Gamma^{(N_m)} \leq \gamma_{u,n}(t) < \infty
            \end{cases}
\end{split}
\end{equation}
where $\left\{\Gamma^{(m)}\right\}_{m=1}^{N_m-1}$, with $\Gamma^{(m)}\leq\Gamma^{(m+1)}$, are the instantaneous SINR boundaries defining the MCS intervals, which can be obtained from \eqref{eq:block-error-rate-relation} as
\begin{equation}
\label{eq:amc-boundaries}
	\Gamma^{(m)}=\frac{1}{\kappa_2^{(m)}}\ln\frac{\kappa_1^{(m)}}{\check{\epsilon}}.
\end{equation}

\paragraph{Continuous-rate allocation (CRA)}: A useful abstraction when exploring rate limits is to assume that each UE's has a set of transmission modes with infinite granularity. In this case, for each value of $\gamma\geq 0$ there is an MCS characterised with a transmission rate
\begin{equation}
   \varrho(\gamma)=\frac{1}{T_o}\log_2\left(1+\frac{\gamma}{\Lambda}\right),
\end{equation}
and a BLER given by
\begin{equation}
   \epsilon\left(\gamma_{u,n}(t),\varrho(\gamma)\right)=\begin{cases}
                                                            1, & \gamma_{u,n}(t)<\gamma \\
                                                            0, & \text{otherwise},
                                                        \end{cases}
\label{eq:ideal-block-error-rate-relation}
\end{equation}
where $\Lambda \geq 1$ represents the coding gap due to the utilisation of a practical (rather than ideal) coding scheme. With $\Lambda=1$ this expression results in the Shannon's capacity limit and allows the comparison of practical AMC-based schemes against fundamental capacity-achieving benchmarks.

Again, given the instantaneous SNR $\gamma_{u,n}(t)$, the transmitter can use \eqref{eq:ideal-block-error-rate-relation} to select the most adequate MCS scheme as the one with transmission rate
\begin{equation}
\label{eq:continuous-rate-amc-ideal-CSIT}
\begin{split}
   \rho\left(\gamma_{u,n}(t)\right) &= \max \left\{ \varrho(\gamma) \,: \,\epsilon \left(\gamma_{u,n}(t),\varrho(\gamma)\right) \leq \check{\epsilon}\right\} \\
                                    &= \frac{1}{T_o}\log_2\left(1+\frac{\gamma_{u,n}(t)}{\Lambda}\right).
\end{split}
\end{equation}

\section{Throughput analysis}\label{sec:Macro_analysis}

Let us define $M_0$ as a positive integer random variable representing the number of UEs in the region served by the tagged BS. Using this definition, the average cell throughput for the downlink of the FFR-aided OFDMA-based cellular network can be expressed as
\begin{equation}\label{e8}
\begin{split}
    \eta = &\sum_{k=1}^{\infty} Pr\{M_0=k\} \sum_{k_C=0}^k {k\choose k_C} \left(P_C\right)^{k_C} \left(1 - P_C\right)^{k-k_C}\\
           &\quad\times \left[|\mathcal{F}_C| \eta_n^C(k_C) + |\mathcal{F}_E| \eta_n^E(k-k_C)\right],
\end{split}
\end{equation}
where $\eta_n^A(k)$ is the average cell throughput on the $n$th RB when there are $k$ UEs in cell area $A$, with $A$ being a token used to represent either the cell-centre region $C$ or the cell-edge region $E$, and $P_C$ denotes the probability that a UE is located in the cell-centre region. As UEs are assumed to be uniformly distributed in the entire cell region, the probability that a UE is located in the cell-centre area is
\begin{equation}\label{e9}
  P_C=\frac{R_{th}^2 - R_{0m}^2}{R_m^2 - R_{0m}^2},
\end{equation}
where $R_{0m}$ denotes the minimum distance of a UE from its serving BS.


Now, defining $M_A$ as a non-negative integer random variable representing the number of UEs in the cell region $A$, the average cell throughput on the $n$th RB allocated to cell region $A$ when $M_A = k$, can be obtained as\footnote{Note that since the channel is assumed to be stationary, from this point onwards the time dependence (i.e., (t)) of
all the variables will be dropped unless otherwise stated.}
\begin{equation}\label{eq:etanA}
  \begin{split}
    &\eta_n^A(k) = \mathbb{E}_{\gamma_n^A|M_A} \left\lbrace \tilde{r}\left(\gamma_n^A\right)|M_A = k \right\rbrace \\
    &\quad= N_{sc} \mathbb{E}_{\gamma_n^A|M_A} \left\lbrace \vartheta\left(\gamma_n^A\right) \rho\left(\gamma_n^A\right)|M_A = k \right\rbrace \\
    &\quad= N_{sc} \mathbb{E}_{\gamma_n^A|M_A} \left\lbrace \left(1-\epsilon\left(\gamma_n^A\right)\right) \rho\left(\gamma_n^A\right)|M_A = k \right\rbrace,
  \end{split}
\end{equation}
where $\gamma_n^A$ denotes the instantaneous SINR experienced on the $n$th RB of cell area $A$.

In the DRA case, using \eqref{eq:dicrete-rate-amc-staircase-ideal-CSIT} and \eqref{eq:dicrete-rate-amc-BLER-staircase-ideal-CSIT} in \eqref{eq:etanA} and expressing the result in bits per second, we obtain
\begin{equation}\label{eq:etanA-DRA}
  \begin{split}
    &\eta_n^A(k) = B_b \sum_{m=1}^{N_m} \varrho^{(m)} \int_{\Gamma^{(m)}}^{\Gamma^{(m+1)}} f_{\gamma_n^A|M_A}(x|k) dx \\
    &\quad-B_b \sum_{m=1}^{N_m} \varrho^{(m)} \int_{\Gamma^{(m)}}^{\Gamma^{(m+1)}} \epsilon(x) f_{\gamma_n^A|M_A}(x|k) dx,
  \end{split}
\end{equation}
where $f_{\gamma_n^A|M_A}(x|k)$ is the probability density function (PDF) of $\gamma_n^A$, conditioned on the event that there are $M_A=k$ UEs in the cell region $A$. Although the second summation in \eqref{eq:etanA-DRA} is analytically tractable and leads to closed form expressions, when using AMC schemes based on powerful Turbo or low-density parity-check (LDPC) codes that are characterised by very steep slopes in the waterfall region of the BLER curves, as is the case in LTE and LTE-A systems, it is a very good approximation (see, for example, \cite{Femenias13,Femenias15robust}) to neglect this term and thus,
\begin{equation}\label{eq:etanA-DRA-approx}
\begin{split}
    \eta_n^A(k) \simeq B_b \sum_{m=1}^{N_m} \varrho^{(m)} \Bigl[&F_{\gamma_n^A|M_A}\left(\Gamma^{(m+1)}|k\right)\\
                                                                   &\quad-F_{\gamma_n^A|M_A}\left(\Gamma^{(m)}|k\right)\Bigr],
\end{split}
\end{equation}
where $F_{\gamma_n^A|M_A}(x|k)$ is the cumulative distribution function (CDF) of $\gamma_n^A$, conditioned on the event that there are $M_A=k$ UEs in the cell region $A$.

In the CRA case, an ideal transmission mode selection is assumed and thus, $\epsilon(\gamma_{u,n}(t))=0$. Hence, using \eqref{eq:continuous-rate-amc-ideal-CSIT} in \eqref{eq:etanA} we have that
\begin{equation}\label{eq:etanA_CRA}
  \begin{split}
    \eta_n^A(k) &= B_b \mathbb{E}_{\gamma_n^A|M_A} \left\lbrace \log_2\left(1 + \frac{\gamma_n^A}{\Lambda}\right)|M_A = k \right\rbrace \\
                &= B_b \log_2 e \int_0^{\infty} \frac{1-F_{\gamma_n^A|M_A}(x|k)}{\Lambda+x} \textup{ d}x.
  \end{split}
\end{equation}

In order to obtain closed-form average throughput expressions, the CDF $F_{\gamma_n^A|M_A}(x|k)$ has to be calculated and this depends on the specific scheduling policy applied by the resource allocation algorithm. In the following subsections, this CDF will be obtained for the PF scheduling rule and then it will be extended to both the MSINR and the RR schedulers.

\subsection{PF scheduling}

As shown in\cite{Choi07}, a proportional fair (PF) scheduler, exploiting the knowledge of the instantaneous SINRs experienced by all UEs $q\in\mathcal{M}_A$, allocates RB $n\in\mathcal{F}_A$ to UE $u\in\mathcal{M}_A$ satisfying
\begin{equation}
   u = \arg\max_{q\in\mathcal{M}_A}\{w_q(t) \gamma_{q,n}(t)\},
   \label{eq:uPF}
\end{equation}
where $\mathcal{M}_A$ is the set indexing all UEs in cell region $A$, and $w_q(t)=1/\mu_q(t)$ is the weighting (prioritisation) coefficient for UE $q$ that, in this case, depends on the short-term averaged evolution of channel-state information, which can be obtained using a moving average over a window of $W$ scheduling periods as
\begin{equation}
   \mu_q(t)=\left(1-\frac{1}{W}\right)\mu_q(t-1)+\sum_{n\in\mathcal{F}_A} \iota_{q,n}(t)\frac{\gamma_{q,n}(t)}{W},
   \label{eq:mu_q}
\end{equation}
with $\iota_{q,n}(t)$ denoting the indicator function of the event that UE $q$ is scheduled to transmit on RB $n$ during scheduling period $t$, that is,
\begin{equation}
  \iota_{q,n}(t)=\begin{cases}
                1, & \text{if UE $q$ is scheduled on carrier $n$ in slot $t$}\\
                0, & \text{otherwise}.
             \end{cases}
\end{equation}

From \eqref{eq:mu_q} we know that, for large values of $W$ and once the PF scheduler reaches stability, $\mu_q(t)$ varies very little with $t$ and thus, it can be safely approximated by its statistical expectation, that is, $\mu_q(t)\simeq\mathbb{E}\{\mu_q(t)\}\triangleq \overline{\mu}_q$. In fact, as stated by Liu and Leung in \cite{Liu10}, experiments suggest that this approximation is valid for $W\geq 50$ with an accuracy greater than $98\%$. Hence, using this approximation and according to the previous definition of the PF scheduler, UE $u \in \mathcal{M}_A$ will be scheduled on RB $n\in\mathcal{F}_A$ whenever
\begin{equation}
   \varphi_{u,n}(t) > \varphi_{\max,u,n}(t) \triangleq \max_{\substack{q\in\mathcal{M}_A \\ q \neq u}} \left\{\varphi_{q,n}(t)\right\},
\end{equation}
where $\varphi_{q,n}(t)\triangleq \gamma_{q,n}(t)/\overline{\mu}_{q}$. That is, UE $u \in \mathcal{M}_A$ is allocated RB $n$ during time slot $t$ if
\begin{equation}
   \gamma_{u,n}(t) > \overline{\mu}_u \varphi_{\max,u,n}(t).
\end{equation}
Thus, since the random variables $\{\varphi_{q,n}(t)\}_{\forall\,q\in\mathcal{M}_A}$ are independent, the conditional CDF of $\gamma_n^A$, conditioned on the event that there are $M_A=k$ UEs in region $A$ and on the set of distances $\boldsymbol{d}=\{d_{0,u}\}_{\forall\,u\in\mathcal{M}_A}$, can be readily evaluated as
\begin{equation}
\begin{split}
   &F_{\gamma_n^A|M_A,\boldsymbol{d}}(\gamma|k,\boldsymbol{d})=\text{Pr}\left\{\gamma_n^A \leq \gamma | M_A=k,\boldsymbol{d}\right\} \\
   &\ =\sum_{\forall\,u\in\mathcal{M}_A} \textrm{Pr}\left\{\gamma_{u,n}(t) \leq \gamma, \varphi_{\max,u,n}(t) \leq \frac{\gamma_{u,n}(t)}{\overline{\mu}_u} \,\Bigr|\,\boldsymbol{d}\right\} \\
   &\ =\sum_{\forall\,u\in\mathcal{M}_A} \int_0^\gamma f_{\gamma_{u,n}|d_{0,u}}(x | d_{0,u}) F_{\varphi_{\max,u,n} | \boldsymbol{d}}\left(\frac{x}{\overline{\mu}_u} \,\Bigr|\,\boldsymbol{d}\right) \mathrm{d} x \\
   &\ =\sum_{\forall\,u\in\mathcal{M}_A} \int_0^\gamma f_{\gamma_{u,n}|d_{0,u}}(x | d_{0,u}) \\
   &\qquad\qquad\qquad\qquad\times \prod_{\substack{q\in\mathcal{M}_A \\ q \neq u}} F_{\varphi_{q,n} | d_{0,q}} \left(\frac{x}{\overline{\mu}_u} \,\Bigr|\,d_{0,q} \right) \mathrm{d} x,
\end{split}
\label{eq:FgammaPF}
\end{equation}
where $F_{\varphi_{\max,q,n} | \boldsymbol{d}}\left(x |\boldsymbol{d}\right)$ is the conditional CDF of $\varphi_{\max,q,n}(t)$ conditioned on the set of distances $\boldsymbol{d}$, and $f_{\gamma_{q,n}|d_{0,q}}(x | d_{0,q})$ and $F_{\varphi_{q,n}|d_{0,q}}(x | d_{0,q})$ are used to denote, respectively, the conditional PDF of $\gamma_{q,n}(t)$ and the conditional CDF of $\varphi_{q,n}(t)$ conditioned on $d_{0,q}$.

In order to obtain further analytical simplifications, it is assumed that, on each RB $n$, the conditional random variables $\{\varphi_{q,n}|d_{0,q}\}_{\forall\,q\in\mathcal{M}_A}$ are independent and identically distributed (i.i.d.). That is, given the positions of UEs in region $A$, it is assumed that on each RB $n$ in region $A$ the UEs are statistically equivalent in terms of the scheduling metrics. As stated by Jingxian Wu \emph{et al.} in \cite{Wu11}, this assumption makes intuitive sense because the fairness of the PF scheduling rule comes indeed from the fact that the scheduling metrics are approximately identically distributed. Using this general assumption and applying integration by parts, the conditional CDF in \eqref{eq:FgammaPF} simplifies to
\begin{equation}
\begin{split}
   &F_{\gamma_n^A|M_A,\boldsymbol{d}}(x|k,\boldsymbol{d}) = \frac{1}{k} \sum_{u\in \mathcal{M}_A} F_{\gamma_{u,n} | d_{0,u}}^k \left(x |d_{0,u} \right).
\end{split}
\end{equation}
Now, taking into account that on each RB $n$ in region $A$, and after averaging over the distance to the BS, the UEs are statistically equivalent in terms of SINR, the (unconditional) random variables $\{\gamma_{q,n}(t)\}_{\forall\,q\in\mathcal{M}_A}$ are i.i.d., and the conditional CDF in \eqref{eq:etanA} can be obtained as
\begin{equation}
    F_{\gamma_n^A|M_A}^{\text{PF}}(x|k)= \int_{R_L^A}^{R_U^A} F_{\gamma_{u,n} | d_{0,u}}^k \left(x |d \right) f_{d_{0,u}}(d) \textrm{d}d,
    \label{eq:FgammanAMA}
\end{equation}
where $R_L^A$ and $R_U^A$ denote the lower and upper radiuses of the circumferences defining cell-region $A$ and $f_{d_{0,u}}(d)$ is the PDF of the random variable $d_{0,u}$ that can be expressed as
\begin{equation}
   f_{d_{0,u}}(d)=\frac{2 d}{{R_U^A}^2-{R_L^A}^2},\ R_L^A \leq d \leq R_U^A.
\end{equation}

The conditional CDF $F_{\gamma_{u,n} | d_{0,u}}(x|d)$ in \eqref{eq:FgammanAMA} can be derived as \cite{Xu12}
\begin{equation}\label{e11}
  \begin{split}
    &F_{\gamma_{u,n} | d_{0,u}}(x|d) = 1-e^{-\frac{x N_0 \Delta f}{\bar{\gamma}_{0,u}}}\\
    &\ + \sum_{b\in\Phi_n} \frac{x\bar{\gamma}_{b,u}^{|\Phi_n|}}{(x\bar{\gamma}_{b,u} + \bar{\gamma}_{0,u})\prod_{\substack{b'\in\Phi_n \\ b'\neq b}} (\bar{\gamma}_{b,u} - \bar{\gamma}_{b',u})} e^{-\frac{x N_0 \Delta f}{\bar{\gamma}_{0,u}}}.
  \end{split}
\end{equation}
where $\overline{\gamma}_{b,u} = P_c L(d_{b,u})$ is the average received power at UE $u$ from BS $b$, and $|\Phi_n|$ is used to denote the cardinality of the set $\Phi_n$.

\subsection{MSINR scheduling}

When implementing the MSINR scheduling rule, in each scheduling period and on each RB $n$ in region $A$, the BS serves the UE experiencing the highest instantaneous SINR, that is,
\begin{equation}
   \gamma_n^A=\max_{q\in\mathcal{M}_A}\left\{\gamma_{q,n}(t)\right\}.
\end{equation}
Note that the MSINR scheduling rule is equivalent to the scheduler specified in \eqref{eq:uPF} by setting the weighting (prioritisation) coefficients to $w_q(t)=1\ \forall\ q\in\mathcal{M}_A$. In this case, following a reasoning similar to that used to analyse the PF scheduling rule, the conditional CDF of $\gamma_n^A$, conditioned on the event that there are $M_A=k$ UEs in region $A$ and on the set $\boldsymbol{d}=\{d_{0,u}\}_{\forall\,u\in\mathcal{M}_A}$, can be expressed as
\begin{equation}
\begin{split}
   &F_{\gamma_n^A|M_A,\boldsymbol{d}}(x|k,\boldsymbol{d})= \prod_{u\in \mathcal{M}_A} F_{\gamma_{u,n} | d_{0,u}} \left(x |d_{0,u}\right).
\end{split}
\end{equation}
Consequently, as on each RB $n$ in region $A$ the UEs are statistically equivalent in terms of SINR, the conditional CDF in \eqref{eq:etanA} simplifies to
\begin{equation}
\begin{split}
    F_{\gamma_n^A|M_A}^{\text{MSINR}}(x|k)
                                          &= \left[\int_{R_L^A}^{R_U^A} F_{\gamma_{u,n} | d_{0,u}} \left(x |d \right) f_{d_{0,u}}(d) \textrm{d}d\right]^k.
\end{split}
\end{equation}

\subsection{RR scheduling}

 A RR scheduler allocates RBs to UEs in a fair time-sharing approach. Since the SINRs experienced by UEs in region $A$ on each RB $n$ are statistically equivalent, serving $M_A=k$ UEs using a RR scheduling policy is equivalent to serving $M_A=1$ UE with MSINR (even when UEs are selected with non uniform probability). Therefore, the conditional CDF in \eqref{eq:etanA} simplifies to
\begin{equation}
\begin{split}
   F_{\gamma_n^A|M_A}^{\text{RR}}(x|k)&=F_{\gamma_n^A|M_A}^{\text{MSINR}}(x|1)\\
                                      &= \int_{R_L^A}^{R_U^A} F_{\gamma_{u,n} | d_{0,u}} \left(x |d \right) f_{d_{0,u}}(d) \textrm{d}d.
\end{split}
\end{equation}

\section{Optimal Designs}\label{sec:Optimisation}

In this section we propose different optimisation designs for the FFR-aided OFDMA-based cellular network aiming at determining the size of the FFR-related spatial and frequency partitions maximising the average cell throughput while satisfying operator-defined system constraints such as using area-proportional frequency partitions or providing a certain fairness degree when allocating data-rates to the cell-centre and cell-edge UEs. In particular, three FFR-based designs are explored:
\begin{itemize}
\item Fixed-spectrum-allocation-factor Design (FxD): In this case the frequency resources are statically and a-priori partitioned between the cell-centre and cell-edge regions. The use of a fixed frequency partition typically provides high spectral efficiencies at the cost of loosing fairness when allocating data rates to cell-centre and cell-edge UEs.
\item Area-proportional Design (ApD): In this design the frequency resources are partitioned proportional to the area of each cell region. Using an area-proportional design allows a more equitable distribution of the spectrum that results in a higher degree of fairness between cell-centre and cell-edge UEs, at the cost of reducing the average cell throughput with respect to the FxD.
\item Quality-constrained Design (QoScD): Under this design the optimisation problem is constrained in order to ensure that the cell-edge throughput is, at least, a fixed fraction of the cell-centre throughput. Consequently, a certain fairness degree (i.e., a certain quality of service (QoS) for the cell-edge UEs) is enforced between cell-centre and cell-edge UEs while maximising the area spectral efficiency under these constraints.
\end{itemize}

The parameters used to pose the optimisation problems are the distance threshold ratio $\omega \triangleq R_{th}/R_m$ and the spectrum allocation factor $\zeta \triangleq N_C/N_b$. It is expected that the appropriate selection of these parameters significantly affects the average cell throughput per RB \cite{Jin13} that, for convenience, is defined as\footnote{Note that, in order to stress its dependency with respect to the optimisation parameters $\zeta$ and $\omega$, the overall average throughput is represented as $\eta(\omega,\zeta)$.}
\begin{equation}\label{d1}
    \tau(\omega,\zeta) \triangleq \frac{\eta(\omega,\zeta)}{B_b N_b} = \zeta \tau^C(\omega,\zeta) + \frac{1-\zeta}{3} \tau^E(\omega,\zeta),
\end{equation}
measured in bps/Hz/RB, where $\tau^A(\omega,\zeta)\triangleq \frac{\eta^A(\omega,\zeta)}{B_b N_A}$ is the average throughput per RB in cell region $A$. Note that $N_C=N_b-3 N_E$ must be a non-negative integer value less or equal than $N_b$ and thus, $N_E\in\{0,1,\ldots,\lfloor N_b/3\rfloor\}$ and the spectrum allocation factor $\zeta$ can only take values in the set \begin{equation}
   \mathcal{S}_\zeta=\left\{\frac{N_b-3\lfloor N_b/3\rfloor}{N_b}, \frac{N_b-3(\lfloor N_b/3\rfloor-1)}{N_b},\ldots,1\right\},
\end{equation}
where $\lfloor x\rfloor$ denotes the floor operator. The optimisation problems described next can be efficiently solved using standard tools typically found in commercial software packages (e.g., Matlab).

\subsection{Fixed-spectrum-allocation-factor Design}

Under FxD, the spectrum allocation factor is fixed to $\zeta = \zeta_o$ (typically $\zeta_o \simeq 0.5$). Therefore, only the parameter $\omega$ remains to be optimised and the problem can be formulated as
\begin{equation}\label{d2}
\omega^* =  \arg\max_{\frac{R_{0m}}{R_m} \leq \omega \leq 1} \zeta_o \tau^C(\omega,\zeta_o) + \frac{1-\zeta_o}{3} \tau^E(\omega,\zeta_o).
\end{equation}

\subsection{Area-proportional Design}
In this case, the spectrum allocation factor is determined by the so-called area-proportional ratio as $\zeta = \omega^2$ \cite{Jin13}. In this case, as $\zeta$ can only take values in the set $\mathcal{S}_\zeta$, the optimisation problem can be formulated as
\begin{equation}\label{d3}
\zeta^* = \arg\max_{\zeta\in\mathcal{S}_\zeta} \zeta \tau^C\left(\sqrt{\zeta},\zeta\right) + \frac{1-\zeta}{3} \tau^E\left(\sqrt\zeta,\zeta\right),
\end{equation}
where, obviously, $\omega^* = \sqrt{\zeta^*}$.

\subsection{QoS-constrained Design}

 In the QoScD approach, a QoS requirement $q$ is stipulated enforcing that the guaranteed average throughput per-RB and per-edge UE is at least a fraction $q$ of the average throughput per-RB and per-centre UE. Hence, the system parameters are adjusted to trade the data rates provided to the cell-centre UEs against those provided to the cell-edge UEs \cite{Jin13}. The corresponding optimisation problem can be formulated as
\begin{equation}\label{d4}
\begin{split}
(\omega^*,\zeta^*) = & \arg\max_{\substack{\frac{R_{0m}}{R_m} \leq \omega \leq 1\\\zeta\in\mathcal{S}_\zeta}} \zeta \tau^C(\omega,\zeta) + \frac{1-\zeta}{3} \tau^E(\omega,\zeta),\\
                    & \textup{ subject to } \tau_{u}^E(\omega,\zeta) \geq q\, \tau_{u}^C(\omega,\zeta),
\end{split}
\end{equation}
where the variables $\tau_{u}^A(\omega,\zeta)$ are defined in a different way for different scheduling policies. Assuming that the average number of UEs served by the BS of interest is equal to $M$ and as the PF and RR schedulers aim at offering either fair throughput values or fair access opportunities to the whole set of UEs, respectively, it seems reasonable to define the per-UE and per-RB average cell throughput as $\tau_{u}^C(\omega,\zeta) = \frac{\zeta \tau^C(\omega,\zeta)}{M P_C}$ for the cell-centre UEs and $\tau_{u}^E(\omega,\zeta) = \frac{(1-\zeta) \tau^E(\omega,\zeta)}{3 M (1-P_C)}$ for the cell-edge UEs. When applying the MSINR scheduling strategy, however, as the scheduler aims at serving the UE experiencing the highest instantaneous SINR, it seems more reasonable to use the definitions $\tau_{u}^C(\omega,\zeta) = \zeta \tau^C(\omega,\zeta)$ and $\tau_{u}^E(\omega,\zeta) = \frac{(1-\zeta) \tau^E(\omega,\zeta)}{3}$ where note that the per-RB average throughput has not been averaged with respect to the average number of UEs in the corresponding cell region.

\begin{table}
\renewcommand{\arraystretch}{1.2}
\caption{Network parameters}
\label{t2}
\centering
\scalebox{0.92}{
\begin{tabular}{|c|c|}
\hline
System parameter & Value\\
\hline
Cell radius & 500 m\\
Minimum distance between BS and UEs & 35 m\\
Antenna configuration & SISO\\
Transmit power of the BS & 46 dBm\\
Antenna gain at the BS & 14 dBi\\
Power spectral density of noise & -174 dBm/Hz\\
Receiver noise figure & 7 dB\\
Total bandwidth & 20 MHz\\
Subcarrier spacing & 15 kHz\\
FFT size & 2048\\
Occupied subcarriers (including DC) & 1201\\
Guard subcarriers & 847\\
Number of resource blocks & 100\\
Subcarriers per RB & 12\\
OFDM symbols per RB (short CP) & 7\\
Useful OFDM symbol duration ($T_o$) & 66.7 {$\mu$}s \\
Short CP duration ($T_{CP}$) & 5.2/4.69 {$\mu$}s \\
Path loss model & $15.3+37.6\log_{10}(d)$ dB\\
\hline
\end{tabular}}
\label{tab:parameters}
\end{table}
\begin{figure*}[!ht]
        \centering
        \begin{subfigure}[MSINR scheduler]{\includegraphics[width=.31\textwidth]{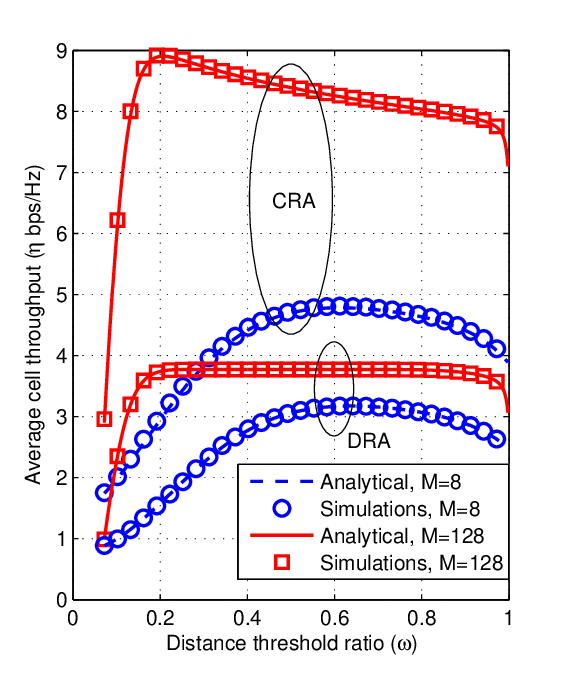}
                \label{fig:FxD_MSINR_Throughput_vs_w}}
        \end{subfigure}
        ~~
        \begin{subfigure}[PF scheduler]{\includegraphics[width=.31\textwidth]{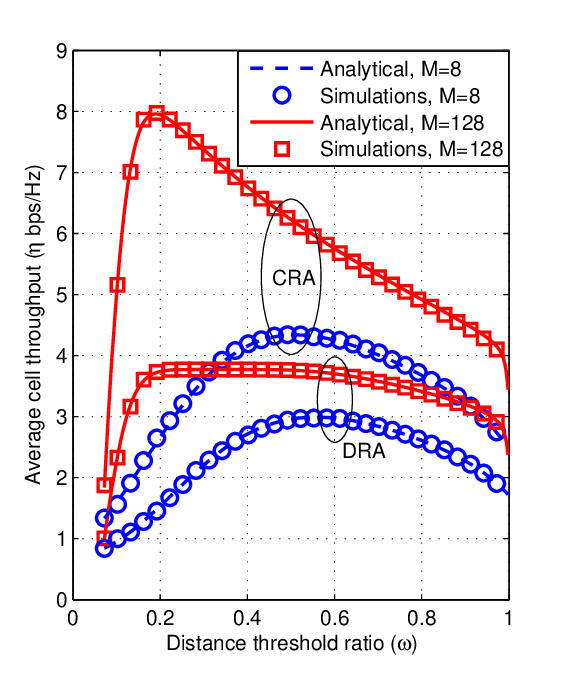}
                \label{fig:FxD_PF_Throughput_vs_w}}
        \end{subfigure}
        ~~
        \begin{subfigure}[RR scheduler]{\includegraphics[width=.31\textwidth]{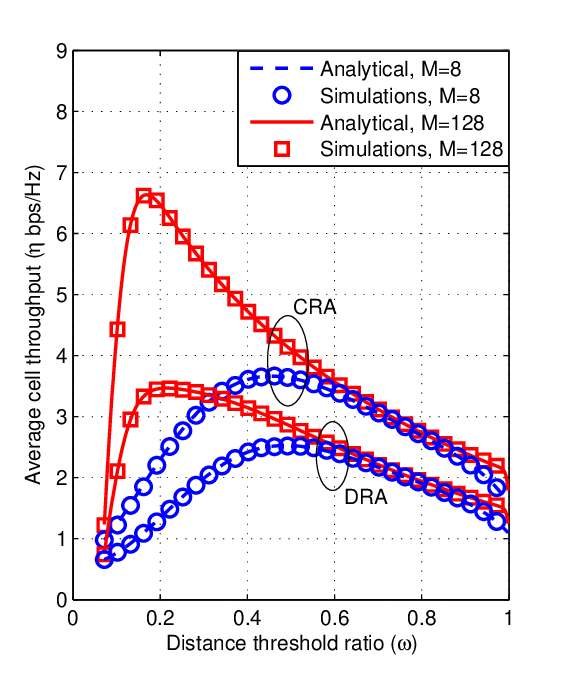}
                \label{fig:FxD_RR_Throughput_vs_w}}
        \end{subfigure}%
        \caption{Average cell throughput versus distance threshold ratio for different values of the average number of UEs per cell $M$ (FxD).}\label{fig:FxD_Throughput_vs_w}
\end{figure*}

\section{Numerical results}\label{sec:Numerical_results}

In order to validate the proposed analytical framework while also providing valuable design guidelines, the 19-cell network shown in Fig. \ref{fig:Network_model} is considered, where the cell of interest is surrounded by two rings of interfering BSs. As stated in previous sections, UEs are distributed over the coverage area using a PPP of normalised intensity $\lambda_m$ (measured in UEs per area unit). For the sake of presentation clarity, although the analytical framework has been developed using the normalised intensity of the PPP, results in this section will be shown as a function of the average number of UEs per cell ($M \triangleq \pi \lambda_m R_m^2$). The main system parameters used to generate both the analytical and simulation results have been particularised using specifications of the downlink of a LTE/LTE-A network \cite{3GPPTS36211,3GPPTS36104,3GPPR9} and have been summarised in Table \ref{tab:parameters}.
\begin{figure*}[!ht]
        \centering
        \begin{subfigure}[Optimal $\omega$]{\includegraphics[width=.31\textwidth]{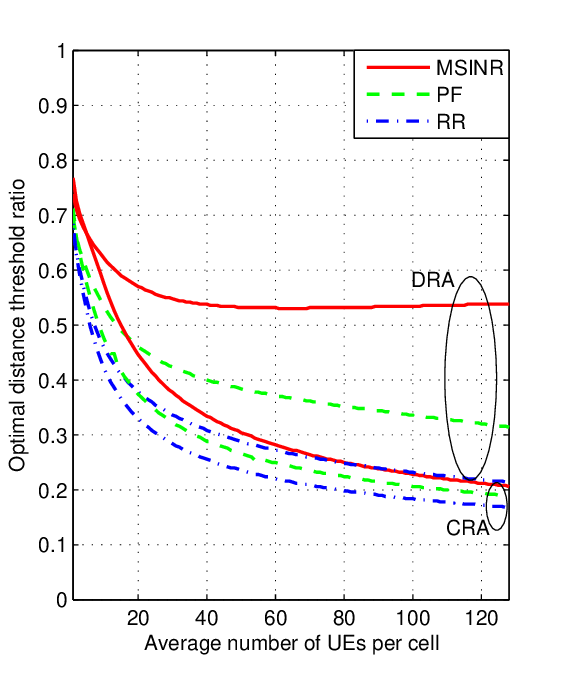}
                \label{fig:FxD_Optimal_w_vs_M}}
        \end{subfigure}
        ~~
        \begin{subfigure}[Optimal throughput]{\includegraphics[width=.31\textwidth]{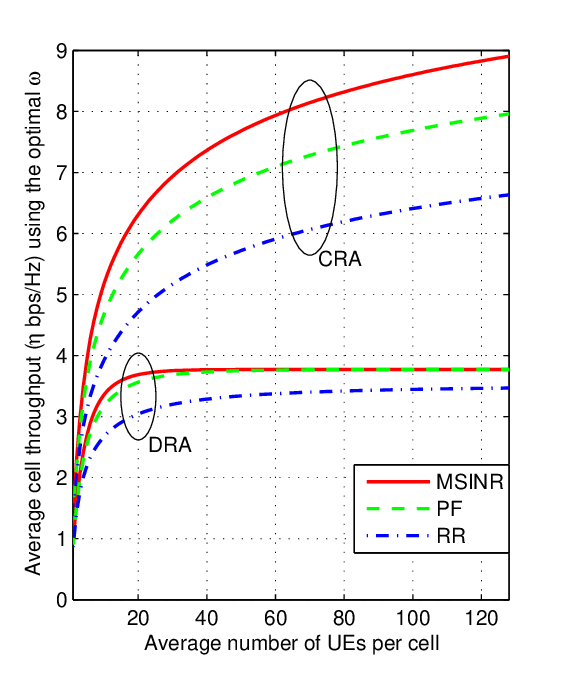}
                \label{fig:FxD_Optimal_Throughput_vs_M}}
        \end{subfigure}
        ~~
        \begin{subfigure}[Cell-region throughput: CRA case.]{\includegraphics[width=.31\textwidth]{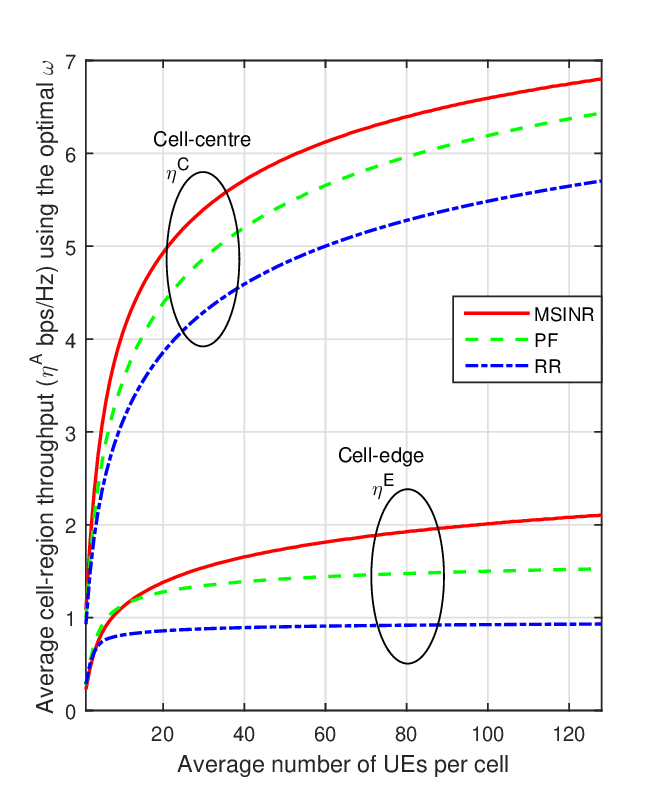}
                \label{fig:Region_Throughput}}
        \end{subfigure}
        \caption{Optimal distance threshold ratio, optimal cell throughput and cell-region throughput versus the average number of UEs per cell $M$ (FxD).}\label{fig:FxD_Optimal_w_and_Throughput_vs_M}
\end{figure*}

\subsection{FxD-based FFR design}

Figure \ref{fig:FxD_Throughput_vs_w} shows the average cell spectral efficiency (measured in bps/Hz) as a function of the distance threshold ratio $\omega$ for the FxD-based FFR design. Illustrating the system behaviour under different network deployments, analytical and simulation results are provided for the MSINR, PF and RR scheduling policies, with the average number of UEs per cell as parameter, and considering the CRA and the DRA strategies. The first point to highlight is the very good match between analytical and simulation results, thus validating the analytical framework developed in Section \ref{sec:Macro_analysis}. Focusing now on performance issues, it can be observed that the average throughput increases with the average number of UEs per cell. This is basically due to two distinct effects. The first one, only exploited by the MSINR and PF scheduling rules, is caused by the larger degree of multiuser diversity provided by the increase of $M$, an effect that is more pronounced for the MSINR scheduler than for the PF scheduling rule because of the tradeoff that the PF scheduler establishes between the exploitation of multiuser diversity and the provision of fairness among UEs. The second effect, affecting all the schedulers but more noticeable when using the RR scheduler, is because increasing the average number of UEs per cell raises the probabilities of having at least one cell-centre UE and one cell-edge UE, hence reducing the probability of ending up with unassigned RBs and the consequent waste of resources. Regarding rate allocation, it is observed that CRA clearly outperforms DRA due to the infinite granularity when mapping SINR to rate, but mostly, because CRA does not have any upper limitation, which in the case of DRA is given by the rate of the highest transmission mode (i.e., SINRs exceeding the highest MCS selection threshold do not bring along any further gains). Note also that a further consequence of this DRA-specific effect is the limited contribution the multiuser diversity gain has on the improvements of the average cell throughput. This is best visualized in the MSINR results, where moving from $M=8$ to $M=128$ does not lead to the dramatic improvement observed when using CRA.

We note that, having established the accuracy of the proposed analytical method, results shown in Fig. \ref{fig:FxD_Optimal_w_and_Throughput_vs_M} depict only the theoretical outcomes, notwithstanding that they have also been duly replicated by means of simulations (not shown here for the sake of clarity). The optimal distance threshold ratio is shown in Fig. \ref{fig:FxD_Optimal_w_vs_M}. Note that, irrespective of the applied scheduling rule and rate allocation strategy, the optimal distance threshold decreases when increasing the average number of UEs per cell, eventually making it rather insensitive to the cell load. This in turn implies that the cell-edge region becomes larger, thus causing the average cell-edge throughput to have a larger impact on the average cell throughput. Again this effect can be largely attributed to the fact that the more users in the system, the lower the probability of having unassigned RBs in the centre region. Notice that, as it can be observed in Fig. \ref{fig:FxD_Optimal_w_vs_M}, irrespective of the rate allocation policy, the fairer the scheduling rule in use the smaller the optimal distance threshold becomes. It is interesting to note in Fig.~\ref{fig:FxD_Optimal_Throughput_vs_M} that regardless of the scheduling technique, CRA is able to exploit the system load increase either because the probability of ending up with empty cell regions diminishes or thanks to the augmented multiuser diversity (for MSINR and PF). In contrast, DRA throughput when using either MSINR or PF tends to quickly saturate at a rather modest cell load basically because the probability of selecting users able to use the highest available MCS rapidly approaches one. In order to illustrate how the overall throughput is distributed between the central and edge regions, Figure \ref{fig:Region_Throughput} shows the average optimal throughputs for each cell-region, centre or edge, as a function of the average number of UEs per cell.

\begin{figure*}[!ht]
        \centering
        \begin{subfigure}[MSINR scheduler]{\includegraphics[width=.31\textwidth]{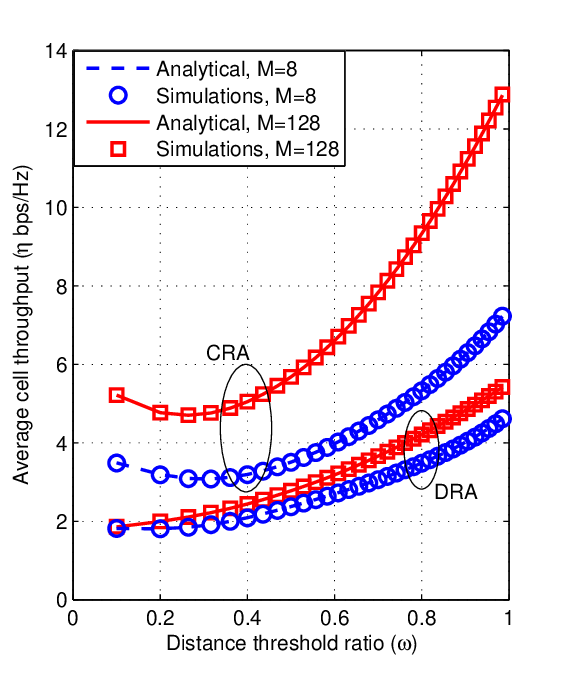}
                \label{fig:ApD_MSINR_Throughput_vs_w}}
        \end{subfigure}
        ~~
        \begin{subfigure}[PF scheduler]{\includegraphics[width=.31\textwidth]{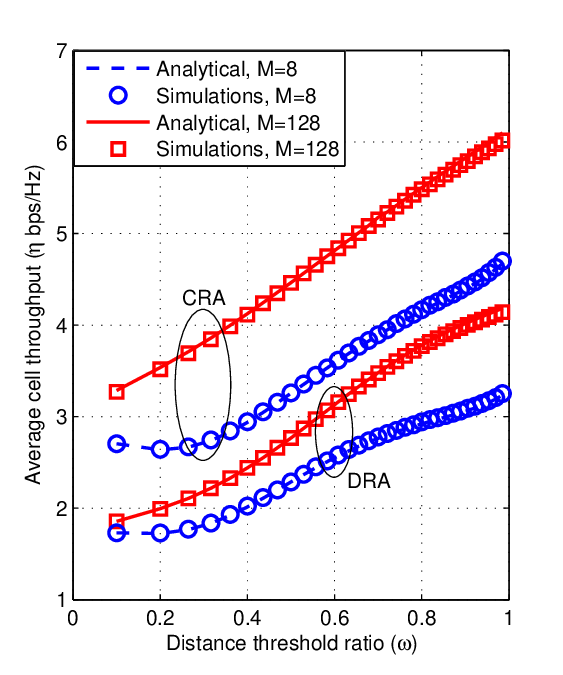}
                \label{fig:ApD_PF_Throughput_vs_w}}
        \end{subfigure}
        ~~
        \begin{subfigure}[RR scheduler]{\includegraphics[width=.31\textwidth]{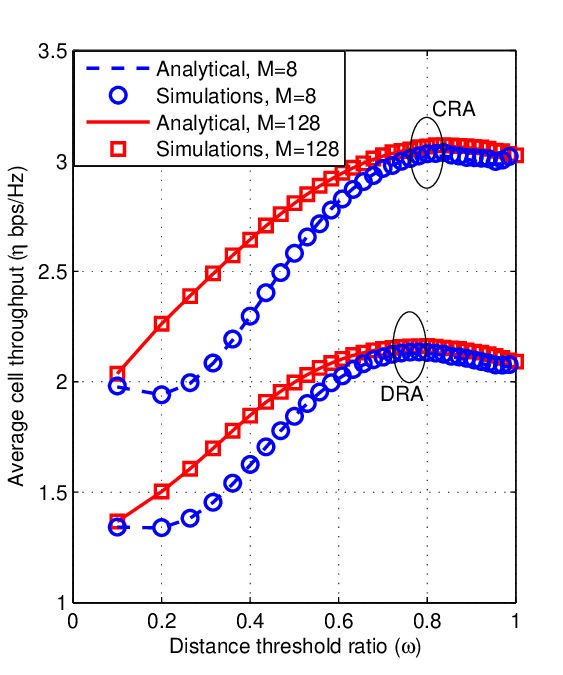}
                \label{fig:ApD_RR_Throughput_vs_w}}
        \end{subfigure}%
        \caption{Average cell throughput versus distance threshold ratio (assuming the use of $\zeta=\omega^2$) for different values of the average number of UEs per cell $M$ (ApD).}\label{fig:ApD_Throughput_vs_w}
\end{figure*}

\subsection{ApD-based FFR design}

In ApD-based optimisation problems the spectrum allocation factor is selected as $\zeta=\omega^2$. Using this setup, Figs. \ref{fig:ApD_MSINR_Throughput_vs_w}, \ref{fig:ApD_PF_Throughput_vs_w} and \ref{fig:ApD_RR_Throughput_vs_w} show, for the MSINR, PF and RR scheduling policies, respectively, the analytical and simulated average cell throughput as a function of the distance threshold ratio $\omega$ and with different configurations of the density of UEs per cell, considering both, the CRA and the DRA strategies. As expected, due to making use of multiuser diversity, the PF and MSINR scheduling rules provide higher average throughput than the RR scheduler. This effect becomes even more evident as the density of UEs per cell increases.

As the MSINR scheduler does not consider the provision of fairness to UEs, it seems quite obvious that the optimal ApD will provide all the available resources to those UEs experiencing the higher SINR values, that is, the cell-centre UEs. This is the result that can be extracted from Fig. \ref{fig:ApD_MSINR_Throughput_vs_w} where, irrespective of the network configuration, the optimal distance threshold ratio is $\omega^*=1$, thus implying that the optimal ApD spectral allocation factor is $\zeta^*=1$. That is, when using MSINR in an ApD-based FFR-aided network, the optimal solution consists of using a full spectrum reuse strategy (cell-edge regions are eliminated and thus, all the RBs are allocated to the cell-centre area). The performance of the PF scheduler, shown in Fig.~\ref{fig:ApD_PF_Throughput_vs_w}, resembles that of the MSINR in the sense that it tends to favour full spectrum reuse but notice that the attained average cell capacity is considerably lower than for the MSINR as some of the multiuser diversity is traded off by an increase in fairness among users. When using the RR scheduling policy, and as it can be observed in Fig. \ref{fig:ApD_RR_Throughput_vs_w}, there is an optimal distance threshold ratio around $\omega=0.8$, that is rather insensitive to the density of UEs in the system. The optimal values of $\omega$ (and hence, those of $\zeta$) are almost independent of $M$ because the RR scheduler does not take advantage of the multiuser diversity.

\begin{figure}[!ht]
        \centering
        \begin{subfigure}{\includegraphics[width=\linewidth]{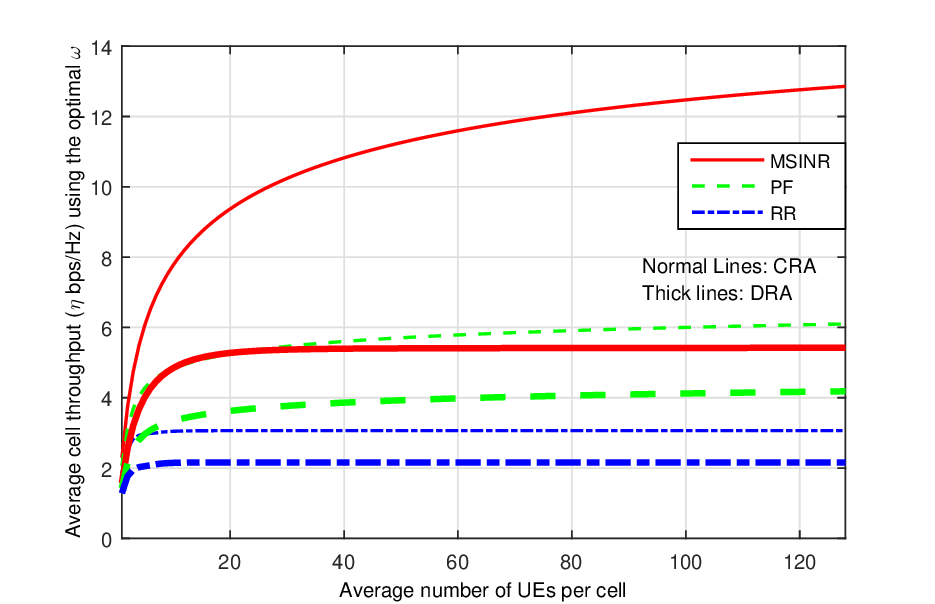}}
        \end{subfigure}
        \caption{Optimal average cell throughput versus the average number of UEs per cell $M$ (ApD).}\label{fig:ApD_Optimal_Throughput_vs_M}
\end{figure}

Figure \ref{fig:ApD_Optimal_Throughput_vs_M} represents, for the CRA and DRA strategies, the average cell throughput using the optimal distance threshold ratio as a function of the  average number of UEs per cell. Notice that in this design, and under MSINR and PF, only the multiuser diversity plays a role in the throughput increase seen with larger cell loads as the optimal distance threshold ratio is $\omega=1$.  In the case of the RR scheduler, the optimal throughput increase stabilises at a lower average number of UEs per cell for both, the CRA and DRA strategies. This is because, as seen in Fig.~\ref{fig:ApD_RR_Throughput_vs_w}, the optimal $\omega$ is rather insensitive to the density of UEs, thus once the probability of ending up with empty cell regions vanishes, there is no further gain in having more users in the system.

\begin{figure*}[!ht]
        \centering
        \begin{subfigure}[MSINR scheduler, $q=0.2$]{\includegraphics[width=.31\textwidth]{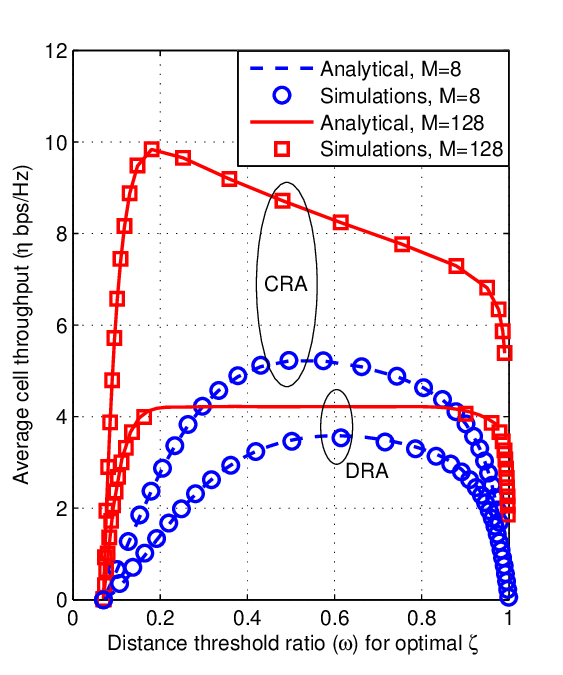}
                \label{fig:QoScD_MSINR_q02_Throughput_vs_w}}
        \end{subfigure}%
        ~~
        \begin{subfigure}[PF scheduler, $q=0.2$]{\includegraphics[width=.31\textwidth]{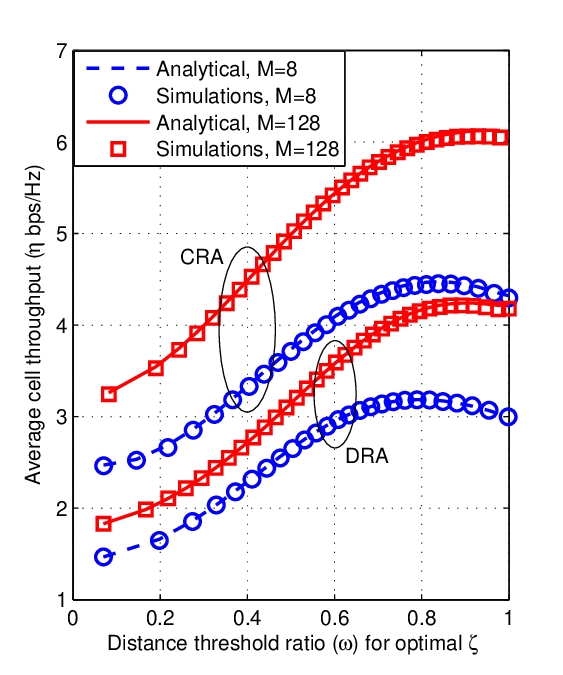}
                \label{fig:QoScD_PF_q02_Throughput_vs_w}}
        \end{subfigure}
        ~~
        \begin{subfigure}[RR scheduler, $q=0.2$]{\includegraphics[width=.31\textwidth]{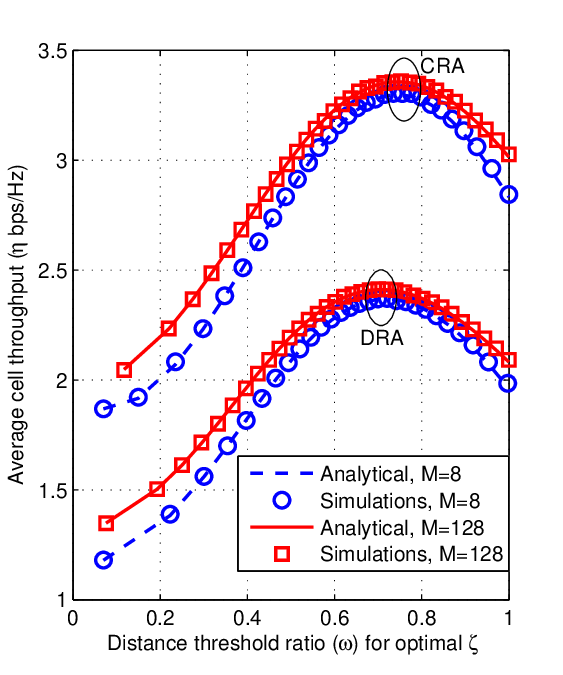}
                \label{fig:QoScD_RR_q02_Throughput_vs_w}}
        \end{subfigure}
        \begin{subfigure}[MSINR scheduler, $q=0.02$]{\includegraphics[width=.31\textwidth]{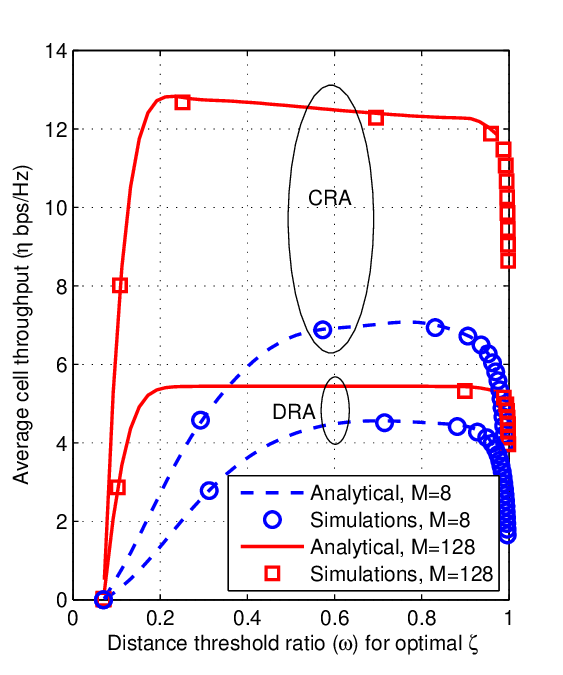}
                \label{fig:QoScD_MSINR_q002_Throughput_vs_w}}
        \end{subfigure}%
        ~~
        \begin{subfigure}[PF scheduler, $q=0.02$]{\includegraphics[width=.31\textwidth]{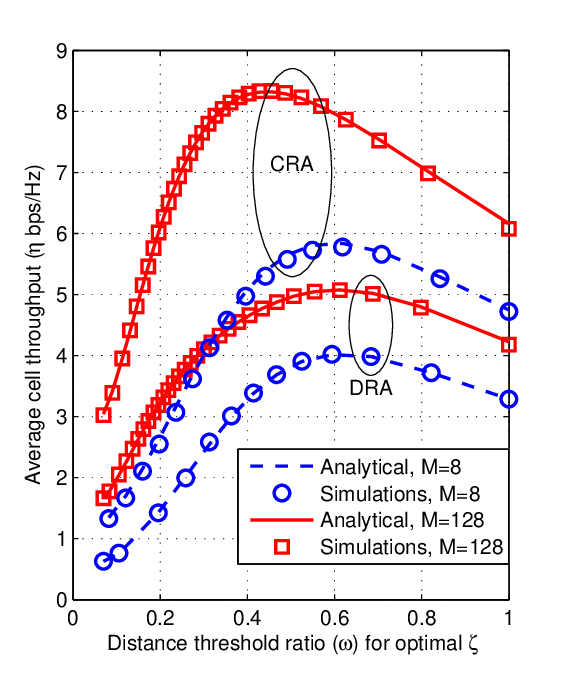}
                \label{fig:QoScD_PF_q002_Throughput_vs_w}}
        \end{subfigure}
        ~~
        \begin{subfigure}[RR scheduler, $q=0.02$]{\includegraphics[width=.31\textwidth]{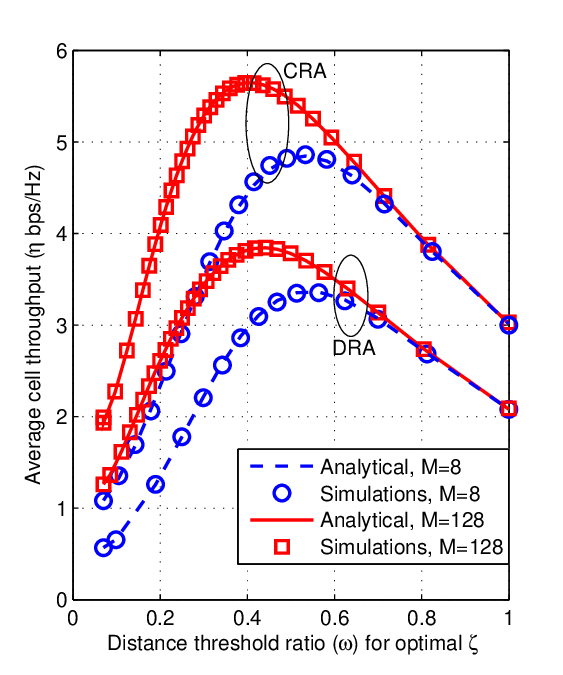}
                \label{fig:QoScD_RR_q002_Throughput_vs_w}}
        \end{subfigure}
        \caption{Average cell throughput versus distance threshold ratio (assuming the use of the corresponding optimal spectrum allocation factor) for different values of the average number of UEs per cell $M$ (QoScD).}\label{fig:QoScD_Throughput_vs_w}
\end{figure*}

\begin{figure*}
        \centering
        \begin{subfigure}[Optimal $\omega$: CRA case.]{\includegraphics[width=.31\textwidth]{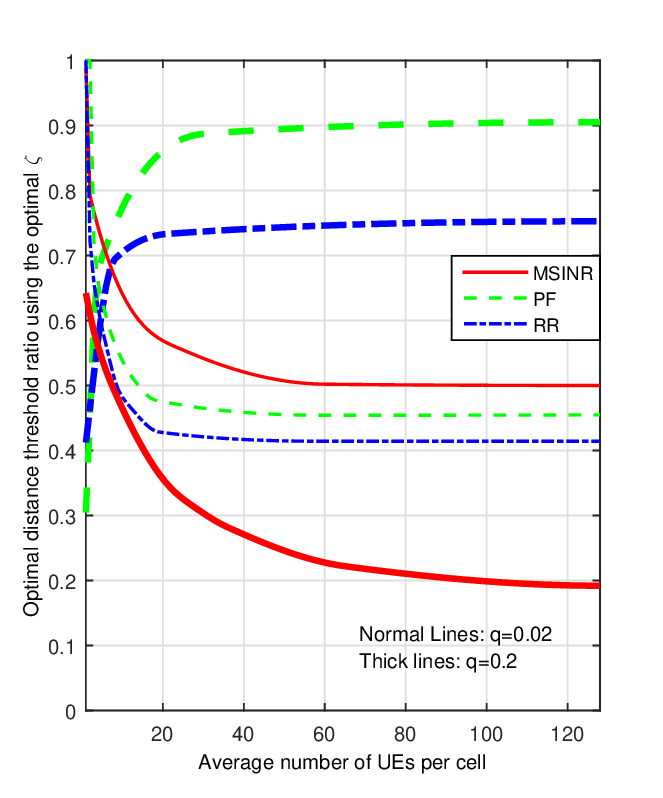}
                \label{fig:QoScD_Optimal_w_vs_M}}
        \end{subfigure}
        ~~
        \begin{subfigure}[Optimal throughput, $q=0.02$]{\includegraphics[width=.31\textwidth]{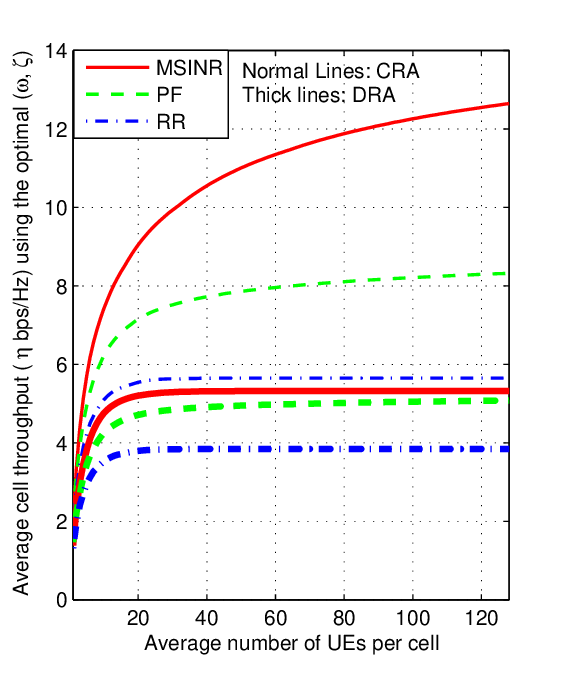}
                \label{fig:QoScD_q002_Optimal_Throughput_vs_M}}
        \end{subfigure}
        ~~
        \begin{subfigure}[Optimal throughput, $q=0.2$]{\includegraphics[width=.31\textwidth]{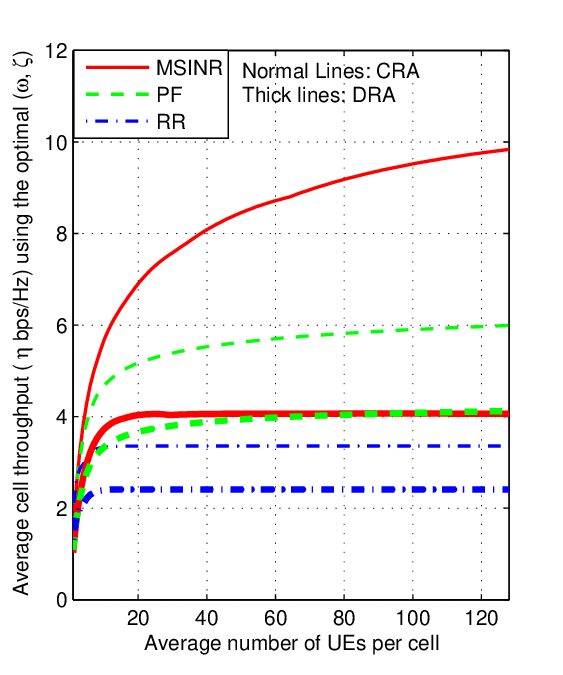}
                \label{fig:QoScD_q02_Optimal_Throughput_vs_M}}
        \end{subfigure}
        \caption{Optimal distance threshold ratio and optimal cell throughput versus the average number of UEs per cell $M$ (QoScD).}\label{fig:QoScD_Optimal_w_and_Throughput_vs_M}
\end{figure*}

\subsection{QoScD-based FFR design}
To assess the performance of the QoScD strategy under different QoS constraints, two different quality factors, $q=0.02$ and $q=0.2$, have been considered, corresponding to low and high throughput fairness requirements between cell-centre and cell-edge UEs. Analytical and simulated average throughput as a function of the distance threshold ratio $\omega$ and with different UE densities are shown in Fig. \ref{fig:QoScD_Throughput_vs_w} for the three considered schedulers. Again, it is worth noting the good agreement between the simulated and analytical results. Also note that for these results, for each value of $\omega$, the value of $\zeta$ maximising the average throughput has been used. The pairs $(\omega,\zeta)$ leading to the maximum average throughput observed in these graphs are indeed the solutions to problem \eqref{d4}.

As in the FxD- and ApD-based FFR designs, the MSINR scheduler provides a clear average throughput performance advantage when compared to the PF and RR schedulers owing to the optimal exploitation of multiuser diversity. Logically, this performance advantage becomes more pronounced in environments with a high density of UEs. Increasing the QoS requirement $q$ enforces a higher degree of fairness between cell-centre and cell-edge UEs at the cost of decreasing the spatially averaged throughput. In ApD-based designs using MSINR scheduling rule we concluded that the full spectrum reuse scheme is the one providing the maximum average system throughput. The full spectrum reuse solution, that is $(\omega,\zeta)=(1,1)$, coincides with that obtained when designing an FFR-aided QoScD-based network with $q=0$. This is obviously a trivial non-realistic QoS constrained design, but it can help us understand the MSINR-related results presented in Figs. \ref{fig:QoScD_MSINR_q02_Throughput_vs_w} and \ref{fig:QoScD_MSINR_q002_Throughput_vs_w}. As the QoScD strategy enforces a minimum degree of fairness between cell-centre and cell-edge UEs, the constrained design aiming at maximising the average system throughput may no longer match the full spectrum reuse scheme, however, the optimal spectrum allocation factor for $q=0.02$ has been found to provide almost 95\% of the available resources to cell-centre UEs, that is, the optimal design tends to the full spectrum reuse strategy. Increasing the QoS requirement from $q=0.02$ to $q=0.2$, produces a notable decrease in both the optimal spectrum allocation factor and the optimal distance threshold ratio, thus increasing the amount of spatial and frequential resources allocated to the cell-edge UEs.

When using the RR scheduler, as it can be observed in Figs. \ref{fig:QoScD_RR_q02_Throughput_vs_w} and \ref{fig:QoScD_RR_q002_Throughput_vs_w}, the optimal distance threshold ratio increases with the quality factor $q$ at the cost of a reduction in the average throughput. In fact, as the QoS constraint is defined as the requirement that $\tau_{u}^E(\omega,\zeta)$ exceeds a certain fraction of $\tau_{u}^C(\omega,\zeta)$, where remember that $\tau_{u}^A(\omega,\zeta)$ is defined as the average per-UE and per-RB throughput in region $A$, strict QoS constraints in an RR-based network can only be achieved by increasing the cell-centre area (i.e., the number of cell-centre UEs) and the number of RBs allocated to the cell-centre UEs. As expected, the performance of the PF scheduler (shown in Figs. \ref{fig:QoScD_PF_q02_Throughput_vs_w} and \ref{fig:QoScD_PF_q002_Throughput_vs_w}) sits in between that of the MSINR and RR, where again it can be observed that increasing $q$ leads to an increase in the optimum distance threshold, and as in the RR case, makes this optimum almost independent of the number of UEs in the system.

Figure \ref{fig:QoScD_Optimal_w_vs_M} shows the optimal distance threshold ratio as a function of the  average number of UEs per cell when using the optimal spectrum allocation factor and for the specific case of CRA. Note that, for the MSINR scheduler and irrespective of the the quality factor, the greater the average number of UEs per cell the lower the optimal distance threshold becomes. This effect is due to the the constraint in \eqref{d4} that causes the cell centre boundary to shrink in an attempt to increase the cell-edge throughput to compensate for the increased probability of selecting central users very close to the BS. For RR and PF schedulers, and for tight QoS requirements (large values of $q$), the optimal distance threshold ratio $\omega$ increases with the average number of users per cell. In contrast, $\omega$ decreases when milder constraints are imposed (low values of $q$). Note that this behaviour can also be observed in Fig.~\ref{fig:QoScD_Throughput_vs_w} and it can be attributed to the effect the constraint in \eqref{d4} has on the per-user throughput performance.
Figures \ref{fig:QoScD_q002_Optimal_Throughput_vs_M} and \ref{fig:QoScD_q02_Optimal_Throughput_vs_M} represent, for $q=0.02$ and $q=0.2$, respectively, the average cell throughput using the optimal distance threshold ratio and the optimal spectrum allocation factor, when considering the CRA and DRA strategies. As expected, using a lower QoS requirement $q$ allows better performance for all scheduling polices and rate allocation strategies. Again, due to the use of a discrete set of MCSs, the DRA strategy (thick lines) achieves lower average cell throughput values than when the CRA strategy is used for the three scheduling polices.

\section{Conclusion}\label{sec:Conclusion}
This paper has introduced and validated an analytical framework to evaluate the performance of FFR-aided OFDMA networks.
Interestingly, the proposed model is able to incorporate the effects of the scheduler and closed-form solutions have been derived for three popular scheduling rules, namely, PF, MSINR and RR. It has been shown that the average system throughput increases with the number of users as the probability of ending up with an empty FFR-defined area (i.e., cell-centre or cell-edge) vanishes, thus avoiding the waste of radio resources. Remarkably, for the specific case of MSINR scheduling, and to a lesser extent for the PF rule, this throughput improvement is further accentuated by a greater exploitation of the multiuser diversity.

Furthermore, the proposed framework allows the incorporation of different rate allocation strategies (CRA and DRA). As anticipated, CRA clearly outperforms DRA in terms of average cell throughput owing to its increased granularity and the avoidance of an upper rate limit that in the case of DRA is defined by the rate of the highest MCS. This effect becomes more significant when employing MSINR scheduling strategies. Regarding the FFR design, it is worth stressing that the optimal threshold radius has been found to significantly depend on the number of users per cell and the scheduling rule implemented at the BS, thus highlighting the importance of analytical models, such as the one proposed in this paper, that allow a quick and accurate performance evaluation under the specific system conditions. Finally, different FFR designs have been considered that suitably dimension the central and edge FFR defined-regions and the amount of frequency resources allocated to each region. In particular, ApD and QoScD designs have been shown to offer different operating points in the throughput \emph{vs} fairness plane.
Further work will concentrate on extending the analysis so as to encompass multi-tier OFDMA heterogeneous networks (e.g. macro and femto tiers) and the use of more sophisticated ICIC techniques (e.g., soft/adaptive frequency reuse, network MIMO).

\section*{Acknowledgements}
Work supported by MINECO (Spanish Government) and FEDER under projects TEC2011-25446 and TEC2014-59255-C3-2-R, and the Conselleria d'Educaci\'o, Cultura i Universitats (Govern de les Illes Balears) under grant FPI/1538/2013 (co-financed by the European Social Fund).

\bibliographystyle{IEEEtran}
\bibliography{MONET-JITEL}

\end{document}